\documentclass[final,12pt,letter]{elsarticle}
\usepackage[]{geometry}
\newgeometry{textwidth=172mm,textheight=222mm}
\usepackage[utf8]{inputenc}
\usepackage{amsmath}
\usepackage{amsfonts}

\date{August 2022}

\journal{Physics Review Letters}

\begin{document}
\begin{frontmatter}

\title{A General Lattice and Basis Formalism Enabling Efficient and Discretized Exploration of Crystallographic Phase Space}

\author[ucb]{David Mrdjenovich}
\author[ucb,lbnl]{Kristin A. Persson\corref{cor1}}

\address[ucb]{Department of Materials Science \& Engineering, 210 Hearst Mining Building, University of California, Berkeley, CA 94720, USA}
\address[lbnl]{Lawrence Berkeley National Laboratory, 1 Cyclotron Road, Berkeley, CA 94720, USA}

\cortext[cor1]{Correspondence: kapersson@lbl.gov}

\begin{abstract}

Three-dimensional lattices are fundamental to solid-state physics. The description of a lattice with an atomic basis constitutes the necessary information to predict solid phase properties and evolution. Here, we present a new algorithm for systematically exploring crystallographic phase space. Coupled with ab-initio techniques, such as density functional theory, this algorithm offers a new approach for exploring and tuning materials behavior, with a broad range of potential applications: polar and magnetic transitions, martensitic phase transformations, and generally materials stability. 

\end{abstract}

\end{frontmatter}

\section{Introduction}
The fundamental description of any crystal consists of a lattice and an atomic basis. This core geometry dictates a myriad of materials properties, informing optical, electronic, mechanical, and chemical functionality.  In materials science and solid-state physics, understanding how the energy varies as a function of its crystallography is crucial to explaining phase transformations and behavior such as piezoelectricity, mechanical stability, ferroelectricity, etc. However, techniques for exploring this energetic landscape, particularly to find low-energy transformation pathways, have remained limited.

The earliest approaches to studying crystallographic transformations involved the calculation of Bain Strains and paths of minimal atomic displacement \citep{bain1924nature}. While these techniques give accurate phase transformation pathways for some materials, they are limited to simple systems and not broadly-generalizable. Most importantly, there is no guarantee that the minimal-displacement path through phase space constitutes the lowest activation energy barrier between the two crystals. Further, due to the fact that strains and cartesian atomic displacements are noncommutative, the definition of minimal-displacement path is, by nature, nebulous for most systems. 

Additional approaches have used Landau theory and symmetry group-subgroup relationships between starting and ending structures to guess possible transformation pathways \citep{stokes2002procedure}. However, these techniques rely on the crystal retaining a certain degree of symmetry throughout the transformation. This is not the case in general, particularly outside the regime of second-order phase transformations. Further, linear interpolation and mapping of Wyckoff labels used in these approaches does not guarantee correspondence with a low-energy pathway.

Other studies have employed specialized solid-state nudged elastic band techniques to search for transformation pathways \citep{sheppard2012generalized}. These approaches show promise for surveying crystallographic phase space more generally; however, ultimately, the search is confined to a small region of phase space dictated by the relaxation of an initial transformation pathway ansatz. It is possible that other transformation pathways exist and that a more-thorough search might identify a lower activation energy barrier minimum in the general case.

Here, we delineate an algorithm for fully exploring crystallographic phase space in a non-redundant, incremental, and rigorous fashion. Further, the approach is discretized, robust to numerical precision errors, and proceeds intuitively. Through the algorithm, we provide new avenues for identifying low-energy transformation pathways, exploring arbitrary paths through phase space, identifying and generating similar crystal structures, and representing crystals numerically. With this work, we aim to simplify the process of exploring and studying how materials transform and how crystal structures are geometrically related, a fundamental part of solid-state physics. 

This letter is organized as follows: in Sec. II we introduce the space of all 3D lattices, their representation as integers, and the calculation of geometrically-similar lattices. In Sec. III we discuss the unique representation of the crystallographic atomic basis as a series of integers. In Sec. IV we combine these techniques, among others, to uniquely represent all possible crystal structures as a string of integers. Finally, we conclude with a summary and an enumeration of future avenues of research.

\section{Representation of 3D Lattices}
Any 3D lattice can be represented by a set of three linearly-independent generating vectors; however, an equivalent and alternative specification exists. Every 3D lattice possesses at least one non-strictly obtuse ``superbasis" $\left\{\vec{v_0}, \vec{v_1}, \vec{v_2}, \vec{v_3}\right\}$, calculable using the Selling Algorithm \citep{conway1992low}. Here $\left\{\vec{v_1}, \vec{v_2}, \vec{v_3}\right\}$ are a generating set for the lattice; $\vec{v_0} := - (\vec{v_1} + \vec{v_2} + \vec{v_3} )$; and the angles between the vectors are all at least $90^\text{o}$. Additionally, the seven square vector lengths $\left\{ v_0^2 , v_1^2 , v_2^2, v_3^2, \left( \vec{v_0} + \vec{v_1} \right)^2, \left( \vec{v_0} + \vec{v_2} \right)^2, \left(\vec{v_0} + \vec{v_3}\right)^2\right\}$, termed here as ``squared vonorms", also represent the lattice. The matrix equation below \citep{conway1992low} relates these vector lengths to the dot products of the superbasis: $\left\{ \vec{v_i} \cdot{} \vec{v_j}\right\}$. Thus, by knowing these seven vector lengths, one may calculate the dot products of an obtuse superbasis of the lattice, thereby reconstructing the lengths and angles of the 3D lattice's generating set with some algebra [eq. 4-11]. 

\begin{equation}\label{eq.1}
    \begin{pmatrix}
-1 & -1 & 1 & 1 & 1 & -1 & -1 \\
-1 & 1 & -1 & 1 & -1 & 1 & -1 \\
-1 & 1 & 1 & -1 & -1 & -1 & 1 \\
1 & -1 & -1 & 1 & -1 & -1 & 1 \\
1 & -1 & 1 & -1 & -1 & 1 & -1 \\
1 & 1 & -1 & -1 & 1 & -1 & -1 \\
-1 & -1 & -1 & -1 & 1 & 1 & 1
\end{pmatrix}
\begin{pmatrix}
v_0^2 \\
v_1^2 \\
v_2^2 \\ 
v_3^2 \\
\left(\vec{v_0} + \vec{v_1}\right)^2 \\
\left(\vec{v_0} + \vec{v_2}\right)^2 \\
\left(\vec{v_0} + \vec{v_3}\right)^2
\end{pmatrix} =
4\begin{pmatrix}
\vec{v_0}\cdot{}\vec{v_1}\\
\vec{v_0}\cdot{}\vec{v_2}\\
\vec{v_0}\cdot{}\vec{v_3}\\
\vec{v_1}\cdot{}\vec{v_2}\\
\vec{v_1}\cdot{}\vec{v_3}\\
\vec{v_2}\cdot{}\vec{v_3}\\
0
\end{pmatrix}
\end{equation}

The set of seven squared vonorms and the aforementioned dot products are a geometric invariant of the lattice: any two lattices related by an affine transformation will have the same set of square vonorms and dot products. By definition, [eq. 1] must be satisfied by all obtuse superbases that exist for a given lattice. However, there may be several distinct obtuse superbases for the same lattice geometry, either related by affine transformations or a re-ordering of the vonorms and dot products.

For the purposes of this and future study, we are concerned only with the effect of crystal geometry on the energy of the system; thus, affine transformations of the crystal can be considered equal in energy. In the presence of external fields, for example, this would not be the case, and orientation degrees of freedom would need to be included in a description of the system.

Studying the re-ordering of distinct superbases, we term the lengths $\left\{ v_0^2, v_1^2, v_2^2, v_3^2\right\}$ ``primary squared vonorms" and the lengths $\left\{ \left( \vec{v_0} + \vec{v_1}\right)^2, \left(\vec{v_0} + \vec{v_2} \right)^2, \left( \vec{v_0} + \vec{v_3} \right)^2 \right\}$ ``secondary squared vonorms". It can be shown by brute force calculation that it is impossible to interchange primary and secondary squared vonorms while preserving the validity of [eq. 1] and the algebraic structure of a superbasis. However, it is possible to permute the labels of the primary squared vonorms arbitrarily. These manipulations constitute the $S_4$ permutation group and signify that there can be up to 24 distinct obtuse superbases representing the same lattice geometry, each superbasis having the same set of lengths and angles in a distinct order. 

Calculating the orbit of the squared vonorms under the group-action of $S_4$ yields a collection of re-ordered primary and secondary squared vonorms. We choose a unique representative (L) from this orbit by choosing the permutation for which the seven ordered lengths, $\left\{ v_0^2 , v_1^2 , v_2^2, v_3^2, \left( \vec{v_0} + \vec{v_1} \right)^2, \left( \vec{v_0} + \vec{v_2} \right)^2, \left(\vec{v_0} + \vec{v_3}\right)^2\right\}$, are maximally-ascending. In a precise sense, we say that for each orbit member (M), there exists an index j such that the first (j - 1) elements of (L) are $\leq$ the first (j - 1) elements of (M) AND the jth element of (L) is strictly less than the jth element of (M). Thus, for any lattice, we can select a unique ordered string of lengths to represent its geometry unambigouously.

As en example, consider the rhombohedral crystal structure of the element, Antimony, where the squared vonorms are as follows (units of $\text{\AA}^2$, rounded to the nearest 0.1) \citep{barrett1963crystal}: 
$$\left\{ v_0^2 = 19.2 ,\  v_1^2 = 21.3,\  v_2^2 = 19.2,\  v_3^2 = 21.3,\  \left( \vec{v_0} + \vec{v_1} \right)^2 = 40.5,\  \left( \vec{v_0} + \vec{v_2} \right)^2 = 19.2,\  \left(\vec{v_0} + \vec{v_3}\right)^2 = 21.3\right\}$$

First we sort the primary square vonorms. This can be accomplished by swapping labels 1 and 2. Notice both secondary square vonorms $(\vec{v_0} + \vec{v_1})^2$ and $(\vec{v_0} + \vec{v_2})^2$ also swap.
$$\left\{ v_0^2 = 19.2 ,\  v_1^2 = 19.2,\  v_2^2 = 21.3,\  v_3^2 = 21.3,\  \left( \vec{v_0} + \vec{v_1} \right)^2 = 19.2,\  \left( \vec{v_0} + \vec{v_2} \right)^2 = 40.5,\  \left(\vec{v_0} + \vec{v_3}\right)^2 = 21.3\right\}$$

Seeing that square vonorms $v_0^2$ and $v_1^2$  are equal, we can swap indices 0 and 1 without changing the order of the primary square vonorms. By swapping 0 and 1, the secondary square vonorms are re-ordered. The result is maximally-ascending:
$$\left\{ v_0^2 = 19.2 ,\  v_1^2 = 19.2,\  v_2^2 = 21.3,\  v_3^2 = 21.3,\  \left( \vec{v_0} + \vec{v_1} \right)^2 = 19.2,\  \left( \vec{v_0} + \vec{v_2} \right)^2 = 21.3,\  \left(\vec{v_0} + \vec{v_3}\right)^2 = 40.5\right\}$$

Then the final unambiguous representation of this lattice is the following string:\\ $\left\{ 19.2, 19.2, 21.3, 21.3, 19.2, 21.3, 40.5\right\}$

From this standardized representation, it is easy to calculate similar lattices. Any two square vonorms out of the seven can be varied by a small amount $\pm \xi$, such that [eq.2] remains true.

\begin{equation}\label{eq.2}
-\left( v_0^2 + v_1^2 + v_2^2 + v_3^2\right) + \left(\vec{v_0} + \vec{v_1}\right)^2 + \left(\vec{v_0} + \vec{v_2}\right)^2 + \left(\vec{v_0} + \vec{v_3}\right)^2 = 0
\end{equation}

This is a necessary invariant of all 3D lattices \citep{conway1992low}. Where $\xi$ is small, the new lengths correspond to a lattice, one that is related to the original by a small strain [proof 1]. Using this co-variation technique, a lattice can have up to 42 different neighboring lattices. Each neighbor will be distinct but geometrically-similar to the original lattice. 

It is possible to use a measuring unit of small size $\xi$. Then, by rounding each vonorm to the nearest $\xi$, while preserving [eq. 2], we can store each lattice as a series of integer coordinates. By varying these coordinates $\pm 1$ it is then possible to calculate the neighbors as before (explicit enumeration in the appendix). Referencing [eq. 1], we observe that, with the given discretization, each dot product must be an integer multiple of $\xi / 4$. By storing the quantities $4 \vec{v_i} \cdot{} \vec{v_j}$, it is possible to represent the dot products also as integers. For any non-strictly obtuse superbasis, each dot product must be less-than or equal-to zero. Thus, out of the 42 total possible lattice neighbors, we keep only those who are distinct with qualifying non-positive dot products. In floating-point arithmetic, due to limited numerical precision, it can be difficult to determine whether a dot product is positive \citep{doi:10.1137/S1064827502407627}. Thus, the decision to use an integer representation has the added bonus of avoiding any ambiguities in this determination. 

It can be shown that, by a series of lattice neighbors, any integer 3D lattice can be connected to any other integer 3D lattice [proof 2]. Further, by design, if lattice A is a neighbor of lattice B, it follows that lattice B is a neighbor of lattice A [proof 3]. Keeping these properties in mind, this discretization scheme creates a grid over the entire configuration space of 3D lattices and provides a straightforward means of exploring the space. In general, neighbors are similar to one another if the discretization parameter is sufficiently small, and two lattices become less similar the more neighbors that separate them.

\section{Representation of Crystalline Atomic Bases}
In the representation of a crystal's atomic basis ($n$ atoms), there are many redundancies in the specification, namely, arbitrary origin, identical atoms, and lattice translations. By choosing relative coordinates in the interval $[0,1)$, arbitrary lattice translations are accounted for. By confining a particular atom to the origin, arbitrary origin is accounted for, and only $n-1$ atomic positions need to be specified. However, where there are identical atoms, there remains possible ambiguity as to which one of the identical atoms is confined to the origin. Further, in an ordered listing of the $n-1$ basis vectors $\left\{ \vec{v_1}, ..., \vec{v}_{n-1}\right\}$, swapping identical non-origin atoms will not change the crystal structure; however, the listing will be re-ordered.

Using the same technique employed with the lattice length strings above, we can calculate the full orbit of strings of basis vectors under their permutation group and then pick a unique representative that is maximally-ascending. Specifically, we order atoms primarily by species and secondarily by first, second, and then third relative coordinate, in ascending numeric order. This uniquely specifies the representation of the crystal basis in terms of a string of numbers.

Finally, by dividing the interval $[0,1)$ into $\delta$ small intervals, it's possible to round each basis coordinate to the nearest interval boundary. Then, the basis can be completely specified as a string of integers, eliminating the possibility of numerical instability and round-off errors.

As an example consider the compound $Sn_2O_4$, a unit cell of the experimentally-observed tin-oxide rutile phase in the canonical setting \citep{yamanaka2000x}. This crystal has the following cell parameters $a = 3.24\text{\AA}$ $b = 4.83\text{\AA}$ $c = 4.83\text{\AA}$ $\alpha = 90^\text{o}$ $\beta = 90^\text{o}$ $\gamma = 90^\text{o}$, when rounded to the nearest hundredth of an angstrom. It also has the following atomic basis, when rounded to the nearest tenth in relative coordinates: \\
2 atoms Sn @0, 0, 0 and @0.5, 0.5, 0.5 \\
4 atoms O @0, 0.7, 0.7 and @0, 0.3, 0.3 and @0.5, 0.8, 0.2 and @0.5, 0.2, 0.8

We decide here to divide the interval $[0,1)$ into 10 small intervals, and we choose atom Sn to be at the origin. Notice that there are two distinct choices for which atom Sn is at the origin.

With the first choice, as written above, the basis can be represented by the following coordinate string: \\
5,5,5, 0,3,3, 0,7,7, 5,2,8 5,8,2

The first Sn atom at the origin is omitted. The second Sn atom is listed first, and its coordinates are discretized. Next, the 4 O atoms are listed. They are sorted by first relative coordinate, then second relative coordinate, then third relative coordinate.

Choosing the second Sn atom to be at the origin, a distinct coordinate string results, following the same ordering protocol: \\
5,5,5, 0,3,7, 0,7,3, 5,3,3 5,8,8

Now comparing the two possible basis strings, element by element, we find that the first string is less than the second string. Thus, the standard basis representation is 5,5,5, 0,3,3, 0,7,7, 5,2,8 5,8,2, the maximally-ascending element of the orbit of equivalent basis atom coordinates, given the discretiztation parameter of 10.

To generate similar crystal atomic bases, each integer relative coordinate of each atomic vector can be individually varied $\pm 1$. Further, we include the vector $\pm\left<1, 1, 1\right>$ as an additional allowed variation (explicit enumeration in the appendix, motivation in [proof 8]). Together these generate $8n$ neighbors for any crystal basis. If $\delta$ is sufficiently large, these neighbors are similar geometrically to the original basis.

\section{Representation of Complete Crystal Structure}
By combining the integer specifications of the lattice and the crystal atomic basis, it is possible to completely specify a crystal using only integers. However, the representation of the crystal atomic basis, being in relative coordinates, requires a particular unit cell of the lattice to be defined. Because each lattice is represented by 7 integers that are in an unambiguous order, the following formulas [eq. 3-10] will always define 3 unambiguous generators for the lattice. Further, this mapping of vonorms to generators is injective: two distinct strings of vonorms leads to two distinct generating sets [proof 4]. Since the crystal's orientation is arbitrary in this study, a variable orientation arises naturally when using these formulas.

\begin{equation}\label{eq.3}
\vec{v_0} = \xi v_0\left<1, 0, 0\right>
\end{equation}
\begin{equation}\label{eq.4}
\vec{v_1} = \xi v_1 \left<x, y, 0\right>
\end{equation}
\begin{equation}\label{eq.5}
x = \frac{\vec{v_0} \cdot \vec{v_1}}{\xi^2v_0v_1}
\end{equation}
\begin{equation}\label{eq.6}
y = \left(1 - x^2\right)^{1/2}
\end{equation}
\begin{equation}\label{eq.7}
\vec{v_2} = \xi v_2 \left< a, b,c\right>
\end{equation}
\begin{equation}\label{eq.8}
a = \frac{\vec{v_0} \cdot \vec{v_2}}{\xi^2v_0v_2}
\end{equation}
\begin{equation}\label{eq.9}
b = \frac{1}{y} \left( \frac{\vec{v_1} \cdot \vec{v_2}}{\xi^2v_1v_2} - xa\right)
\end{equation}
\begin{equation}\label{eq.10}
c = \left( 1 - a^2 - b^2\right)^{1/2}
\end{equation}

For conciseness, we group these relations together in a function $\mathbb{L} : \mathbb{Z}^7 \rightarrow \mathbb{R}^{3\times 3}$. $\mathbb{L}$ takes in the 7 integer vonorms as a vector, ordered as $v_0^2$, $v_1^2$, $v_2^2$, $v_3^2$, $(\vec{v_0} + \vec{v_1})^2$, $(\vec{v_0} + \vec{v_2})^2$, $(\vec{v_0} + \vec{v_3})^2$. $\mathbb{L}$ uses these 7 values to calculate the corresponding dot products, using [eq. 1]. Finally, $\mathbb{L}$ outputs the three generating vectors as column vectors of a 3x3 matrix ordered as follows: $[\vec{v_0}, \vec{v_1}, \vec{v_2}]$

Now we recall the symmetry group $S_4$, which can permute the order of square vonorms but leaves the geometry of the lattice unchanged. In the context of the function $\mathbb{L}$, each $g \in S_4$ can lead to a distinct unit cell and orientation. In general, $\mathbb{L}(g(\vec{vo})) = A_g\mathbb{L}(\vec{vo})\mu_g^{-1}$, where $A_g$ is a pure rotation or rotoinversion (unitary), and $\mu_g$ is a unimodular matrix [proof 5]. Considering this change in unit cell, the relative basis coordinates must be left-multiplied by $\mu_g$ to account for the change, whereas, with the affine transformation, $A$, the relative basis coordinates do not change [proof 6].

For completeness, we list a generating set of $S_4$ and their unimodular counterparts $\mu_g$. There is an isomorphism between these groups [proof 7]; thus, any permutation and corresponding matrix can be constructed by the analogous group multiplication. We restrict $\mu_g$ to have a positive determinant so that only proper rotations are required to connect neighboring crystal structures [proof 5]. Physically, we base this restriction on the idea that when connecting points in crystallographic phase space, basis atoms are never allowed to invert through the origin. Such an inversion is physically prohibited, as atoms would approach infinitesimally close to one another. Further, inversion of the basis typically represents a large, abrupt change in crystal structure. Our goal here is to consider small crystallographic perturbations, composing them together to naturally reconstruct larger structural transformations.

\begin{equation}\label{eq.11}
0 \leftrightarrow 1: \mu_g = -\begin{pmatrix}0 & 1 & 0 \\ 1 & 0 & 0 \\ 0 & 0 & 1\end{pmatrix}
\end{equation}
\begin{equation}\label{eq.12}
0 \leftrightarrow 2: \mu_g = -\begin{pmatrix}0 & 0 & 1 \\ 0 & 1 & 0 \\ 1 & 0 & 0\end{pmatrix}
\end{equation}
\begin{equation}\label{eq.13}
0 \leftrightarrow 3: \mu_g = -\begin{pmatrix}-1 & 0 & 0 \\ -1 & 1 & 0 \\ -1 & 0 & 1\end{pmatrix}
\end{equation}
\begin{equation}\label{eq.14}
1 \leftrightarrow 2: \mu_g = -\begin{pmatrix}1 & 0 & 0 \\ 0 & 0 & 1 \\ 0 & 1 & 0\end{pmatrix}
\end{equation}
\begin{equation}\label{eq.15}
1 \leftrightarrow 3: \mu_g = -\begin{pmatrix}1 & -1 & 0 \\ 0 & -1 & 0 \\ 0 & -1 & 1\end{pmatrix}
\end{equation}
\begin{equation}\label{eq.16}
2 \leftrightarrow 3: \mu_g = -\begin{pmatrix}1 & 0 & -1 \\ 0 & 1 & -1 \\ 0 & 0 & -1\end{pmatrix}
\end{equation}

Previously, we considered the orbit of vonorms under the action of $S_4$; however, in the context of a lattice + crystal basis, we need to consider the simultaneous orbit of the atomic basis relative coordinates as well. Here, we define $\tilde{\vec{vo}}$ as the maximally-ascending representative of the $S_4$ orbit on the string of lattice lengths. Some elements in $S_4$ will leave $\tilde{\vec{vo}}$ unchanged. Together these form a subgroup $\mathbb{H}$ of $S_4$ (orbit-stabilizer theorem).  We can orbit the basis coordinates with all the elements of $\mathbb{H}$, forming the set of all possible geometrially-equivalent crystal bases for the given unit cell $\mathbb{L}(\tilde{\vec{vo}})$. For each one of these geometrically-equivalent bases, we calculate the corresponding maximally-ascending basis coordinate string. Then, comparing these strings to one another, we choose a representative that is globally maximally-ascending. Together, the maximally-ascending lattice vonorms and this maximally-ascending basis string constitute a unique representation of the crystal as a string of numbers, termed here ``crystal normal form".

We note that the decision to use integers as a representation removes ambiguity from the calculation of $\mathbb{H}$. If floating-point arithmetic were used, the determination as to which vonorms are equal in value would be difficult and ultimately would rely on a tolerance. Further, because integers are used, we can guarantee that high-symmetry lattices, those with many equal vonorms, will be included and well-represented in the discretization of crystallographic phase space.

Taken together, the aforementioned sets of 42 vonorm co-variations alongside the $8n$ crystal basis variations create a set of geometrically-similar crystals, termed here as crystal neighbors (example in the appendix). By decreasing $\xi$ and increasing $\delta$ these variations can be made arbitrarily small. When a neighbor is calculated, the vonorms or basis relative coordinates are varied, then, the neighbor is put into crystal normal form. This ensures that each crystal is canonically-represented and calculated only one time, eliminating redundancies. Under this scheme, it can be shown that neighbor relationships between crystal structures are reciprocal and that any crystal structure can be connected to any other via a series of neighbor connections, aka. small structural perturbations [proof 8].

As a final note, there are many additional ways that lattice and crystal basis neighbors could be defined. However, given the completeness of this scheme, any alternative neighbors defined can be expressed as a series of one or more aforementioned neighbors, termed here “canonical neighbors”. In general, we expect the canonical neighbors to correspond to the smallest possible strains and phonon mode amplitudes for a given $\xi$ and $\delta$ discretization. Thus, adding additional neighbors will effectively correspond to “larger” step sizes that duplicate the action of multiple canonical neighbors.

\section{Summary}

As presented, any crystal structure can be connected to any other crystal structure with the same formula unit via a series of similar neighbors: small incremental strains and phonon mode amplitudes. These neighbor relationships are reciprocal, forming a well-defined grid. Further each crystal structure can be uniquely represented as a series of integers. As such, the entire 3D crystallographic configuration space, for a fixed number of atoms per unit cell, can be discretized and explored using the parameters $\xi$ and $\delta$.

As an example, we present a canonical transformation pathway from HCP to BCC found using these techniques in Figure 1. Here two Zirconium atoms per unit cell are searched in the configuration space. We pick a shortest path connecting the structures, signifying a path with the smallest number of nearest-neighbor connections.

\begin{figure}
    \centering
    \includegraphics[width=\textwidth]{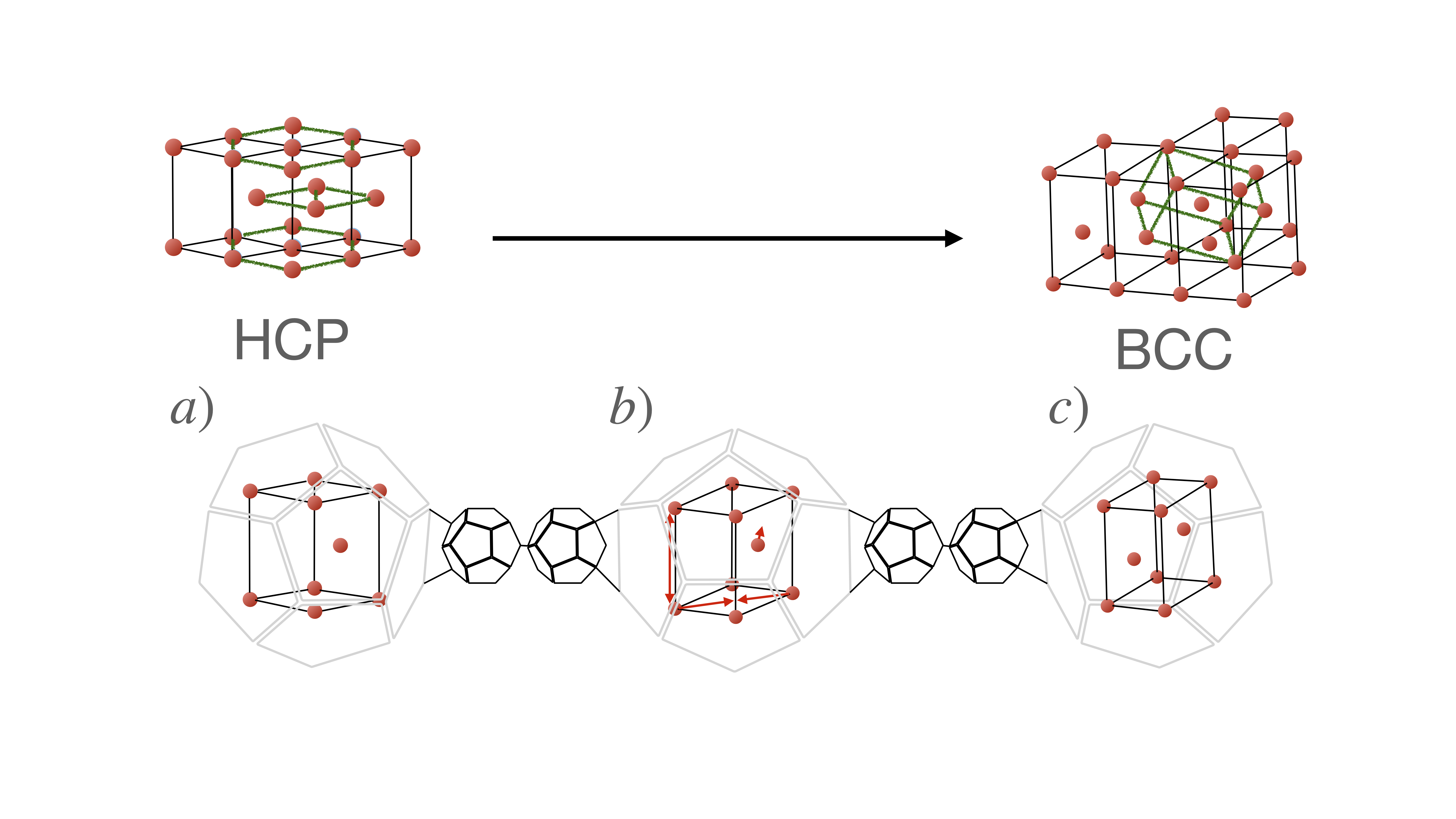}
    \caption{The search for transformation pathways from HCP to BCC is visualized below. The hexagonal closest-packed planes are highlighted in green for the HCP structure. The canonical BCC unit cell is highlighted in green for the BCC structure. The unique unit cells defined by canonical mapping function $\mathbb{L}(\vec{vo})$ are shown for HCP (a) and BCC (c). An intermediate structure connects the two (b) with unit cell strain and atomic shuffling visualized via red arrows. Some neighbors are omitted for conceptual clarity. The path highlighted is a shortest path connecting the HCP and BCC crystals. This path closely corresponds to the well-known C' strain, TA1(N) phonon-mode transformation pathway connecting BCC and HCP structures in many metallic systems \citep{ekman1998ab}}.
    \label{fig}
\end{figure}

Employing this approach, we anticipate many possible applications to the field of solid-state physics and materials science. By studying a general collection of crystal structures and evaluating their energies using first-principles techniques, we can gain greater insight into the phase space of materials and assess materials stability, among other functional properties. The larger energetic landscape reveals activation energy barriers separating distinct phases of a material and can elucidate general transformation pathways induced by external fields, such as magnetic, electric and mechanical energy gradients. The curvature around energetic minima yields second order response functions, such as elastic tensors and various phonon frequencies. Ultimately, an energy landscape generated using this technique may also enable a new way to sample phase space using Monte-Carlo approaches to estimate finite temperature properties using ab-initio techniques. While the orientation of the crystal system has been explicitly discounted in this treatment, with additional effort, the strain and rotation connecting neighboring crystal structures can be calculated using techniques such as \citep{koumatos2016optimality}. With these parameters, it is possible to keep track of the overall crystal orientation relative to an external applied field or similar. In this way, coupled properties, such as structural variation in response to external fields or changes in spin state as a response to structural perturbation can be assessed under a consistent global coordinate system.

From a materials engineering perspective, by using this representation in combination with a dataset of materials energies, we anticipate possible utility in fitting interatomic potentials and training machine learning algorithms. Using an entire energy landscape as a training input, would add rich structure-energy information outside of the well-known crystal structures and their perturbations, likely improving the generality of fitted models. Additionally, this technique offers another way to assess crystallographic similarity between materials: the fewer nearest neighbor hops required, the more similar the crystals. This may prove useful in large databases of crystal structures to quickly screen for duplicates in the database.

\section{Acknowledgements}
This work gratefully acknowledges support from  the U.S. Department
of Energy, Office of Science, Office of Basic Energy Sciences, Materials Sciences
and Engineering Division, for their support under Contract No. DE-AC02-
05-CH11231 within the Data Science for Data-Driven Synthesis Science grant
(KCD2S2). D.M. would like to sincerely thank Dr. Kristin Persson for her unwavering support and guidance through the duration of this project.

\section{Proof 1: Nearest neighbor lattices are connected by small strains}

Assume there exist 2 lattices, $\mathbb{L}_1$ and $\mathbb{L}_2$, with similar square vonorms, differing by a small amount $\sim \xi$.

Assume there exists a transformation $A$ such that $[\mathbb{L}_2] = A[\mathbb{L}_1]$, where $[\mathbb{L}]$ is a 3x3 matrix with 3 generating vectors of $\mathbb{L}$ arranged as column vectors.

Then, it follows: $\left<A \vec{v_i}, A\vec{v_i}\right> = \left< \vec{v_i}, \vec{v_i}\right> + \xi_i$, where $\vec{v_i}$ corresponds to one of the seven voronoi vectors, and $\xi_i$ is the difference in the square vonorm value between lattices $\mathbb{L}_1$ and $\mathbb{L}_2$ for the $i^{th}$ square vonorm.

Re-arranging: $\left<\vec{v_i}, A^TA\vec{v_i}\right> - \left< \vec{v_i}, \vec{v_i}\right> = \xi_i$, using the standard definitions of matrix multiplication and conventional cartesian inner product on $\mathbb{R}^n$.

Re-arranging: $\xi_i = v_i^2 \left<\hat{v_i}, \left(A^TA - I\right)\hat{v_i}\right>$, using linearity and the distributive property of the inner product, where $\hat{v_i}$ represents a unit vector in the direction of $\vec{v_i}$. 

Assume $A^TA - I$ has an orthonormal eigenbasis $\{\hat{b_i}\}$ with eigenvalues $\{\lambda_i\}$, noting that $A^TA - I$ is real and symmetric, thereby invoking the spectral decomposition theorem.

We can express $\hat{v_i}$ as a linear combination of eigenbasis vectors: $\hat{v_i} = \sum_j c_j \hat{b_j}$

Re-arranging: $\frac{\xi_i}{v_i^2} = \sum_{j,k} \left< c_j\hat{b_j}, \lambda_k c_k \hat{b_k}\right> = \sum_j c_j^2 \lambda_j$

Further, note that: $\sum_i c_i^2 = 1$, as $\hat{v_i}$ and $\hat{b_i}$ are all unit vectors.

Thus, $\xi_i / v_i^2$ is a non-negative weighted average of all $\lambda_i$ values. When the set of $\xi_i / v_i^2$ values are near-zero for all $i$, it must follow that for many different weightings, the weighted average of all $\lambda_i$ values is small. Thus, each $\lambda_i$ value must be near-zero.

Invoking the polar decomposition theorem, $A = RT$, where $R$ is unitary and $T$ is symmetric. Physically, $R$ corresponds to a rotoinversion and $T$ corresponds to a stretch. Thus, the quantity, $A^TA - I = T^2 - I$, and the eigenvalues, $\lambda_i = s^2 - 1$, correspond to a square principal stretch value minus 1.

Thus, where $\lambda_i$ values are small, it follows that the principal stretch values must be near 1, meaning a small strain connects the lattices.

\section{Proof 2: Every lattice can be connected to every other lattice via a series of nearest neighbors}

We start by referencing [proof 3]: every lattice neighbor relationship is reciprocal. Thus, to show lattices $\mathbb{L}_1$ and $\mathbb{L}_2$ are connected by a series of neighbors, it suffices to show that they are both connected to an intermediate lattice $\tilde{\mathbb{L}}$.

With a bit of algebra, it can be shown that the following matrix equation applies for any superbasis, whether or not obtuse:

$$\frac{1}{2}\begin{pmatrix}-1&-1&0&0&1&0\\-1&0&-1&0&0&1\\0&1&1&0&-1&-1\\1&0&0&1&-1&-1\\0&-1&0&-1&0&1\\0&0&-1&-1&1&0\end{pmatrix}
\begin{pmatrix}
v_0^2 \\ v_1^2 \\ v_2^2 \\ v_3^2 \\ (\vec{v_0} + \vec{v_1})^2 \\ (\vec{v_0} + \vec{v_2})^2
\end{pmatrix}
=
\begin{pmatrix}
\vec{v_0} \cdot{} \vec{v_1} \\
\vec{v_0} \cdot{} \vec{v_2} \\
\vec{v_0} \cdot{} \vec{v_3} \\
\vec{v_1} \cdot{} \vec{v_2} \\
\vec{v_1} \cdot{} \vec{v_3} \\
\vec{v_2} \cdot{} \vec{v_3}
\end{pmatrix}$$

Considering a subset of vonorm modifications $\{\vec{\chi_i}\}$, listed completely in the appendix, we can calculate the image of each modification when multiplied by the above matrix. Notice only the first 6 square vonorms are listed: the 7th is uniquely implied by [eq. 2].

$$\{\vec{\chi_i} \} = \left\{\begin{pmatrix}0\\0\\1\\0\\1\\0\end{pmatrix},\begin{pmatrix}0\\0\\0\\1\\1\\0\end{pmatrix},\begin{pmatrix}0\\1\\0\\0\\0\\1\end{pmatrix},\begin{pmatrix}0\\0\\0\\1\\0\\1\end{pmatrix},\begin{pmatrix}0\\1\\0\\0\\0\\0\end{pmatrix},\begin{pmatrix}0\\0\\1\\0\\0\\0\end{pmatrix},\begin{pmatrix}1\\0\\0\\0\\0\\0\end{pmatrix},\begin{pmatrix}0\\0\\0\\1\\0\\0\end{pmatrix},\begin{pmatrix}1\\0\\0\\0\\0\\1\end{pmatrix},\begin{pmatrix}0\\0\\1\\0\\0\\1\end{pmatrix},\begin{pmatrix}1\\0\\0\\0\\1\\0\end{pmatrix},\begin{pmatrix}0\\1\\0\\0\\1\\0\end{pmatrix}\right\}$$

$$A\{\vec{\chi_i}\} = \frac{1}{2}\left\{\begin{pmatrix}1\\-1\\0\\-1\\0\\0\end{pmatrix},\begin{pmatrix}1\\0\\-1\\0\\-1\\0\end{pmatrix},\begin{pmatrix}-1\\1\\0\\-1\\0\\0\end{pmatrix},\begin{pmatrix}0\\1\\-1\\0\\0\\-1\end{pmatrix},\begin{pmatrix}-1\\0\\1\\0\\-1\\0\end{pmatrix},\begin{pmatrix}0\\-1\\1\\0\\0\\-1\end{pmatrix},\begin{pmatrix}-1\\-1\\0\\1\\0\\0\end{pmatrix},\begin{pmatrix}0\\0\\0\\1\\-1\\-1\end{pmatrix},\begin{pmatrix}-1\\0\\-1\\0\\1\\0\end{pmatrix},\begin{pmatrix}0\\0\\0\\-1\\1\\-1\end{pmatrix},\begin{pmatrix}0\\-1\\-1\\0\\0\\1\end{pmatrix},\begin{pmatrix}0\\0\\0\\-1\\-1\\1\end{pmatrix}\right\}$$

Noting that the matrix above is invertible, it can be seen that if all dot products $\vec{v_i} \cdot{} \vec{v_j}$ are zero, then all vonorms must be zero. As such, every 3D lattice must have at least one dot product strictly less than zero.

Examining the images of $\{\vec{\chi_i}\}$ above, every 3D lattice has at least one neighbor that keeps all dot products non-positive. Further, by combining multiple of these neighbors together in a series, the dot products $\vec{v_i} \cdot{} \vec{v_j}$ can be made arbitrarily negative. Here we call this specific summation and series of neighbors a canonical obtuse step. Further, when derived in this order, the neighbors, relative to their full $S_4$ orbit, remain maximally-ascending at each step. This is because larger vonorms are incremented first, before smaller vonorms are later incremented by the same amount. Thus, the ordering is always unchanged.

$\vec{\chi_i} + \vec{\chi_j} ... = \begin{pmatrix}0\\0\\0\\1\\0\\0\end{pmatrix}+\begin{pmatrix}0\\0\\1\\0\\0\\1\end{pmatrix}+\begin{pmatrix}0\\1\\0\\0\\1\\0\end{pmatrix}+\begin{pmatrix}0\\0\\0\\1\\0\\0\end{pmatrix}+\begin{pmatrix}0\\0\\1\\0\\0\\1\end{pmatrix}+\begin{pmatrix}1\\0\\0\\0\\1\\0\end{pmatrix}+\begin{pmatrix}0\\0\\0\\1\\0\\0\end{pmatrix}+\begin{pmatrix}0\\1\\0\\0\\0\\1\end{pmatrix}+\begin{pmatrix}1\\0\\0\\0\\1\\0\end{pmatrix}+\begin{pmatrix}0\\0\\1\\0\\0\\0\end{pmatrix}+\begin{pmatrix}0\\1\\0\\0\\0\\1\end{pmatrix}+\begin{pmatrix}1\\0\\0\\0\\1\\0\end{pmatrix}$

$A(\vec{\chi_i} + \vec{\chi_j} ...) = \frac{1}{2}\left(\begin{pmatrix}0\\0\\0\\1\\-1\\-1\end{pmatrix}+\begin{pmatrix}0\\0\\0\\-1\\1\\-1\end{pmatrix}+\begin{pmatrix}0\\0\\0\\-1\\-1\\1\end{pmatrix}+\begin{pmatrix}0\\0\\0\\1\\-1\\-1\end{pmatrix}+\begin{pmatrix}0\\0\\0\\-1\\1\\-1\end{pmatrix}+\begin{pmatrix}0\\-1\\-1\\0\\0\\1\end{pmatrix}+\begin{pmatrix}0\\0\\0\\1\\-1\\-1\end{pmatrix}+\begin{pmatrix}-1\\1\\0\\-1\\0\\0\end{pmatrix}+\begin{pmatrix}0\\-1\\-1\\0\\0\\1\end{pmatrix}+\begin{pmatrix}0\\-1\\1\\0\\0\\-1\end{pmatrix}+\begin{pmatrix}-1\\1\\0\\-1\\0\\0\end{pmatrix}+\begin{pmatrix}0\\-1\\-1\\0\\0\\1\end{pmatrix}\right) = \begin{pmatrix}

-1 \\ -1 \\ -1 \\ -1 \\ -1 \\ -1

\end{pmatrix}$

With these preliminaries, we begin the proof.

Take two lattices $\mathbb{L}_1$ and $\mathbb{L}_2$. Begin by adding first any necessary $\vec{\chi_i}$ values from above so that all $\vec{v_i}\cdot{}\vec{v_j}$ values are at most -3/2. We will then start with these derived neighbors $\tilde{\mathbb{L}}_1$ and $\tilde{\mathbb{L}}_2$, ensuring that the vonorms are in maximally-ascending order. At any time during the course of the steps below, if any dot products of either lattice become -1 or larger, both lattices must be modified by a canonical obtuse step to reduce the dot products. This ensures that specific neighbors can continue to be derived while keeping the superbasis obtuse. Applying a canonical obtuse step to both lattices concurrently will not change the difference between the vonorms of both lattices.

Using $\vec{\chi_i} = \left<0, 0, 0, 1, 0, 0, 1 \right>$, we construct consecutive neighbors from the lattice with the smaller $v_3^2$ square vonorm until both lattices have equal values for $v_3^2$.

Using $\vec{\chi_i} = \left<0, 0, 1, 0, 0, 0, 1 \right>$, we construct consecutive neighbors from the lattice with the smaller $v_2^2$ square vonorm until both lattices have equal values for $v_2^2$.

Using $\vec{\chi_i} = \left<0, 1, 0, 0, 0, 0, 1 \right>$, we construct consecutive neighbors from the lattice with the smaller $v_1^2$ square vonorm until both lattices have equal values for $v_1^2$.

Using $\vec{\chi_i} = \left<1, 0, 0, 0, 0, 0, 1 \right>$, we construct consecutive neighbors from the lattice with the smaller $v_0^2$ square vonorm until both lattices have equal values for $v_0^2$.

At this point, both original lattices $\mathbb{L}_1$ and $\mathbb{L}_2$ have been connected via nearest neighbors to new lattices $\tilde{\mathbb{L}}_1'$ and $\tilde{\mathbb{L}}_2'$. Both of these new lattices have identical primary square vonorm values. Further, the progression above ensures that the primary square vonorms are always sorted in ascending order at each intermediate step. Thus, no permutation changing the order of primary square vonorms occurs.

Using $\vec{\chi_i} = \left<0, 0, 0, 0, \bar{}1, 0, 1 \right>$, we construct consecutive neighbors from the lattice with the smaller $(\vec{v_0} + \vec{v_3})^2$ square vonorm until both lattices have equal values for $(\vec{v_0} + \vec{v_3})^2$.

Using $\vec{\chi_i} = \left<0, 0, 0, 0, \bar{}1, 1, 0 \right>$, we construct consecutive neighbors from the lattice with the smaller $(\vec{v_0} + \vec{v_2})^2$ square vonorm until both lattices have equal values for $(\vec{v_0} + \vec{v_2})^2$.

At this point, both original lattices $\mathbb{L}_1$ and $\mathbb{L}_2$ have been connected via nearest neighbors to new lattices $\tilde{\mathbb{L}}_1''$ and $\tilde{\mathbb{L}}_2''$. Both of these lattices have identical primary square vonorm values and identical values for $(\vec{v_0} + \vec{v_3})^2$ and $(\vec{v_0} + \vec{v_2})^2$. Further, the progression above ensures that the secondary square vonorms are always sorted in maximally-ascending order at each intermediate step. Thus, no permutation changing the order of secondary square vonorms occurs.

Finally, referring to [eq. 2], the value of 6 square vonorms determines the value of the remaining vonorm. In this case, because both lattices $\tilde{\mathbb{L}}_1''$ and $\tilde{\mathbb{L}}_2''$ have the same 6 square vonorms, they must have the same 7 square vonorms. Thus, both lattices must be equal. As such, any two lattices $\mathbb{L}_1$ and $\mathbb{L}_2$ can be connected by a series of neighbors to a common lattice. Thus, any two lattices can be connected to one another via nearest neighbors.

\section{Proof 3: Lattice neighbor relationshps are reciprocal:}

Assume lattice $\mathbb{L}_A$ has neighbor $\mathbb{L}_B$. Mathematically this means that when a displacement $\vec{\chi_i}$ is added to the vonorms of $\mathbb{L}_A$ and the vonorms are sorted to be maximally-ascending, the vonorms of $\mathbb{L}_B$ result.

$$g ( \vec{vo}_A + \vec{\chi_i}) = \vec{vo}_B : g \in S_4$$

We can re-arrange this equation algebraically:

$$g^{-1}(\vec{vo}_B - g(\vec{\chi_i})) = \vec{vo}_A$$

Referencing the appendix where $\{\vec{\chi_i}\}$ are listed, it can be seen that every permutation $g \in S_4$ is a bijection on this set. That is, by construction, the set is closed under this group action. Further for each $\vec{\chi_i}$ in the set, its negative is included.

Thus, the action $-g$ maps $\vec{\chi_i}$ onto another element of the set: $\vec{\chi_j}$

$$-g(\vec{\chi_i}) = \vec{\chi_j}$$

Thus, it can be said that there exists a displacement such that: 

$$g(\vec{vo}_B + \vec{\chi_j}) = \vec{vo}_A : g \in S_4$$

As such, lattice $\mathbb{L}_B$ has neighbor lattice $\mathbb{L}_A$ by definition.

\section{Proof 4: The lattice mapping function $\mathbb{L}(\vec{vo})$ is injective}

For this proof, we reference the following truncated relationship between vector norms and dot products for any superbasis:

$$\frac{1}{2}\begin{pmatrix}-1&-1&0&0&1&0\\-1&0&-1&0&0&1\\0&1&1&0&-1&-1\\1&0&0&1&-1&-1\\0&-1&0&-1&0&1\\0&0&-1&-1&1&0\end{pmatrix}
\begin{pmatrix}
v_0^2 \\ v_1^2 \\ v_2^2 \\ v_3^2 \\ (\vec{v_0} + \vec{v_1})^2 \\ (\vec{v_0} + \vec{v_2})^2
\end{pmatrix}
=
\begin{pmatrix}
\vec{v_0} \cdot{} \vec{v_1} \\
\vec{v_0} \cdot{} \vec{v_2} \\
\vec{v_0} \cdot{} \vec{v_3} \\
\vec{v_1} \cdot{} \vec{v_2} \\
\vec{v_1} \cdot{} \vec{v_3} \\
\vec{v_2} \cdot{} \vec{v_3}
\end{pmatrix}$$

Assume that $\mathbb{L}(\vec{vo}_1) = \mathbb{L}(\vec{vo}_2)$. Then, the generating vectors output must be identical from the function, and they must have equal lengths and angles.

Noting equal generator lengths, it follows that $v_0^2$, $v_1^2$, and $v_2^2$ must be equal for both input square vonorm vectors $\vec{vo}_1$ and $\vec{vo}_2$.

Noting equal angles, the dot products $\vec{v_0} \cdot{} \vec{v_1}$, $\vec{v_0} \cdot{} \vec{v_2}$, and $\vec{v_1}\cdot{}\vec{v_2}$ must be equal for both input square vonorm vectors $\vec{vo}_1$ and $\vec{vo}_2$.

From above, $\vec{v_0} \cdot{} \vec{v_1} = -v_0^2 - v_1^2 + (\vec{v_0} + \vec{v_1})^2$. Because $v_0^2$ and $v_1^2$ are equal for both $\vec{vo}_1$ and $\vec{vo}_2$, it must follow that $(\vec{v_0} + \vec{v_1})^2$ is equal for both $\vec{vo}_1$ and $\vec{vo}_2$.

From above, $\vec{v_0} \cdot{} \vec{v_2} = -v_0^2 - v_2^2 + (\vec{v_0} + \vec{v_2})^2$. Because $v_0^2$ and $v_2^2$ are equal for both $\vec{vo}_1$ and $\vec{vo}_2$, it must follow that $(\vec{v_0} + \vec{v_2})^2$ is equal for both $\vec{vo}_1$ and $\vec{vo}_2$.

From above, $\vec{v_1} \cdot{} \vec{v_2} = v_0^2 + v_3^2 - (\vec{v_0} + \vec{v_1})^2 - (\vec{v_0} + \vec{v_2})^2$. Because $v_0^2$, $(\vec{v_0} + \vec{v_1})^2$, and $(\vec{v_0} + \vec{v_2})^2$ are equal for both $\vec{vo}_1$ and $\vec{vo}_2$, it must follow that $v_3^2$ is equal for both $\vec{vo}_1$ and $\vec{vo}_2$.

Combining these results with [eq. 2], because 6 square vonorms are equal, all 7 square vonorms must be equal between $\vec{vo}_1$ and $\vec{vo}_2$. Thus, $\vec{vo}_1 = \vec{vo}_2$.

\section{Proof 5: The group $S_4$ has an action on set of lattice generators $\mathbb{L}(\vec{vo})$ arranged as column vectors. This action is as follows: $\mathbb{L}(g(\vec{vo})) = A_g\mathbb{L}(\vec{vo})\mu_g^{-1} : g \in S_4$, where $A_g$ is a unitary rotoinversion and $\mu_g$ is a unimodular matrix.}

To show a valid group action, we algebraically note the following, referencing that unitary and unimodular matrices form groups under composition:

$$\mathbb{L}(g_1 \cdot{} g_2 ( \vec{vo})) = A_{g_1 \cdot{} g_2} \mathbb{L}(\vec{vo})\mu_{g_1 \cdot{} g_2}^{-1} = A_{g_1} A_{g_2} \mathbb{L}(\vec{vo}) \mu_{g_2}^{-1}\mu_{g_1}^{-1} = \mathbb{L}(g_1(g_2(\vec{vo})))$$

$$\mathbb{L}(e(\vec{vo})) = A_e^{-1}\mathbb{L}(\vec{vo})\mu_e = \mathbb{L}(\vec{vo})$$

As a justification for this definition of group action, consider the action of $S_4$ on the set of seven square vonorms. The lengths and angles of the superbasis are preserved at all times, just re-ordered. The geometry of the lattice is preserved; although, the mapping function $\mathbb{L}(\vec{vo})$ may change the orientation of the lattice. This is represented by a left multiplication by a unitary matrix, which preserves lengths and angles. Again, considering the preservation of lattice geometry, the unit cell $\mathbb{L}(\vec{vo})$ must geometrically be a unit cell of the original lattice. Thus, right multiplication by a unimodular matrix represents this geometric constraint.

Further, $S_4$ is simply a permutation of the superbasis vectors labels, and it’s the closure of the following simple transpositions: $0 \leftrightarrow 1$, $0 \leftrightarrow 2$, $0 \leftrightarrow 3$, $1 \leftrightarrow 2$, $1 \leftrightarrow 3$, $2 \leftrightarrow 3$. If any transposition occurs, the unit cell $\mathbb{L}(\vec{vo})$, defined geometrically by the generators $\vec{v_0}$, $\vec{v_1}$, and $\vec{v_2}$, must change to reflect the transposition.

In the order, $0 \leftrightarrow 1$, $0 \leftrightarrow 2$, $0 \leftrightarrow 3$, $1 \leftrightarrow 2$, $1 \leftrightarrow 3$, $2 \leftrightarrow 3$, the generating set $\{\vec{v_0}, \vec{v_1}, \vec{v_2}\}$ gets orbited to the following: $\{\vec{v_1}, \vec{v_0}, \vec{v_2}\}$, $\{\vec{v_2}, \vec{v_1}, \vec{v_0}\}$, $\{-\vec{v_0} - \vec{v_1} - \vec{v_2}, \vec{v_1}, \vec{v_2}\}$, $\{\vec{v_0}, \vec{v_2}, \vec{v_1}\}$, $\{\vec{v_0}, -\vec{v_0} - \vec{v_1} - \vec{v_2}, \vec{v_2}\}$, $\{\vec{v_0}, \vec{v_1}, -\vec{v_0} - \vec{v_1} - \vec{v_2}\}$, noting that $\vec{v_3} = -(\vec{v_0} + \vec{v_1} + \vec{v_2})$.

By inspection, these transpositions can be accomplished by the right multiplication of the following unimodular matrices.

$$0 \leftrightarrow 1: \mu_g = \pm\begin{pmatrix}0 & 1 & 0 \\ 1 & 0 & 0 \\ 0 & 0 & 1\end{pmatrix}$$

$$0 \leftrightarrow 2: \mu_g = \pm\begin{pmatrix}0 & 0 & 1 \\ 0 & 1 & 0 \\ 1 & 0 & 0\end{pmatrix}$$

$$0 \leftrightarrow 3: \mu_g = \pm\begin{pmatrix}-1 & 0 & 0 \\ -1 & 1 & 0 \\ -1 & 0 & 1\end{pmatrix}$$

$$1 \leftrightarrow 2: \mu_g = \pm\begin{pmatrix}1 & 0 & 0 \\ 0 & 0 & 1 \\ 0 & 1 & 0\end{pmatrix}$$

$$1 \leftrightarrow 3: \mu_g = \pm\begin{pmatrix}1 & -1 & 0 \\ 0 & -1 & 0 \\ 0 & -1 & 1\end{pmatrix}$$

$$2 \leftrightarrow 3: \mu_g = \pm\begin{pmatrix}1 & 0 & -1 \\ 0 & 1 & -1 \\ 0 & 0 & -1\end{pmatrix}$$

In listing the unimodular matrices, $\pm$ is included to represent a sign ambiguity, as all lattices have inversion symmetry in their group structure. Therefore, a unit cell or its inversion will have the same geometry and order of square vonorms. However, when we consider the presence of a crystal basis, we need to pick the appropriate sign for which no inversion occurs. Physically, this ensures that each neighbor involves only a small change in basis. Recall that inversion is a large discontinuous change in crystal basis in the general case: when the basis does not also share inversion symmetry.

Re-arranging the definition of the group action:

$$A_g = \mathbb{L}(g(\vec{vo}))\mu_g\mathbb{L}(\vec{vo})^{-1}$$

We can take the determinant of both sides:

$$|A_g| = |\mathbb{L}(g(\vec{vo}))||\mu_g||\mathbb{L}(\vec{vo})^{-1}|$$

We note that, as defined, $\mathbb{L}(\vec{vo})$ always outputs a right-handed set of generators and has a positive determinant.

Thus it follows that:

$$\text{sign}|A_g| = \text{sign}|\mu_g|$$

The constraint that no inversion occur signifies that $|A_g| > 0$, thus, $|\mu_g| > 0$.

As such, we pick the negative representatives above so that each $\mu_g$ has a positive determinant. The set of unimodular matrices with positive determinant likewise forms a group, so the conclusions above are unchanged.

\section{Proof 6: A unitary rotoinversion of the crystal does not change the relative coordinates of the crystal basis; however, a unimodular change of unit cell does change the relative coordinates of the crystal basis.}

Consider a crystal with basis $\mathbb{B}$, represented as a matrix with each basis atom’s cartesian coordinates corresponding to a column vector. Assume $\mathbb{L}$ to be a matrix representing the 3 generators of a lattice, arranged as column vectors of the matrix. Then we define: $[\mathbb{B}]_\mathbb{L} := \mathbb{L}^{-1}\mathbb{B}$, where $[\mathbb{B}]_\mathbb{L}$ is a matrix with each basis atom’s relative coordinates corresponding to a column vector.

When a unitary rotoinversion $A$ of the crystal occurs, $\mathbb{L} \rightarrow A \mathbb{L}$, $\mathbb{B} \rightarrow A \mathbb{B}$. That is, the lattice generators and cartesian basis coordinates are all modified correspondingly. However, calculating the relative basis coordinates, we see they are unchanged:

$$[\mathbb{B}]_\mathbb{L} \rightarrow \mathbb{L}^{-1}A^{-1}A\mathbb{B} = \mathbb{L}^{-1}\mathbb{B} = [\mathbb{B}]_\mathbb{L} $$

When a unimodular matrix $\mu_g$ changes the unit cell of the crystal, no geometric change has occurred; thus, the cartesian basis coordinates remain the same. However, the geometry of the lattice generators has changed: $\mathbb{L} \rightarrow \mathbb{L}\mu_g$. Calculating the relative basis coordinates, we see they are changed:

$$[\mathbb{B}]_\mathbb{L} \rightarrow \mu_g^{-1}\mathbb{L}^{-1}\mathbb{B} = \mu_g^{-1}[\mathbb{B}]_\mathbb{L}$$

\section{Proof 7: There exists an isomorphism between $S_4$ and the group of unimodular matrices accomplishing the change of unit cell $\mu_g$}

Referencing the group action from [proof 5], the homomorphism from $S_4$ to $\{\mu_g\}$ is already defined through the generator relations.

Assume that $\mu_g = I$. This implies that $g \in S_4$ does not change the order of $\vec{v_0}$, $\vec{v_1}$, and $\vec{v_2}$. Thus, $g$ can only orbit $\vec{v_3}$ with itself and must leave the other primary voronoi vectors unchanged. Thus, $g$ can only be the identity permutation, and the homomorphism has a trivial kernel. This means the homomorphism is injective.

We take $\{\mu_g\}$ to be the set of all products of the generators defined in [proof 5]. As such, for every $\mu_g$ there exists $g \in S_4$ that maps to $\mu_g$ by the homomorphism. This means the homomorphism is surjective.

Taken together, this means that the homomorphism is an isomorphism (bijection) between $S_4$ and the set $\{\mu_g\}$.

\section{Proof 8: The structure of crystal neighbors is reciprocal and complete.}

If crystal B is a neighbor of crystal A, one of the following must be true:

a) If B is a lattice neighbor: \\
$\vec{vo}_B = g(\vec{vo}_A + \vec{\chi_i}) \qquad \mathbb{B}_B = \mu_g\mathbb{B}_A \qquad g \in S_4$

b) If B is a crystal basis neighbor: \\
$\vec{vo}_B = \vec{vo}_A \qquad \mathbb{B}_B = \mu_g (\mathbb{B}_A + \vec{\delta_a}) \qquad g \in S_4$

We can re-arrange the equations as before:

a) If B is a lattice nighbor: \\
$\vec{vo}_A = g^{-1}(\vec{vo}_B - g(\vec{\chi_i})) \qquad \mathbb{B}_A = \mu_g^{-1}\mathbb{B}_B \qquad g \in S_4$

b) If B is a crystal basis neighbor: \\
$\vec{vo}_B = \vec{vo}_A \qquad \mathbb{B}_A = \mu_g ^{-1}(\mathbb{B}_B - \mu_g\vec{\delta_a}) \qquad g \in S_4$

Inspecting both cases: \\

a) $\{\vec{\chi_i}\}$ is closed under $S_4$ permutation and negation: $-g(\vec{\chi_i}) = \vec{\chi_j}$. Thus, if B is a lattice neighbor of A, A will be a lattice neighbor of B. \\

b) It can be shown that the set $\{\vec{\delta_i}\}$ in the appendix is invariant under the action of the generators of $\{\mu_g\}$ and negation: $-\mu_g \vec{\delta_i} = \vec{\delta_j}$. Thus, if B is a crystal basis neighbor of A, A will be a crystal basis neighbor of B.

Therefore, neighbor relationships are reciprocal.

Continuing the proof, we reference [proof 2] to demonstrate that any crystal lattice can be reached as a series of lattice neighbors from any other crystal lattice. Thus, we need only show that for a fixed unit cell reference frame (implying a fixed lattice), all sets of integer basis coordinates can be reached as a series of crystal neighbors. Together this would imply that any combination of lattice and basis can be reached via a series of neighbors.

To demonstrate this, we take two bases within the same unit cell $\mathbb{B}_1$ and $\mathbb{B}_2$. Due to the integer specification and the fact that the displacements $\{\vec{\delta_i}\}$ span the vector space of all atomic displacements, there exists a finite series of displacements (aka. basis coordinate neighbors) connecting the two bases: $\mathbb{B}_2 = \mathbb{B}_1 + \sum \vec{\delta_i}$

Now, throughout the algorithm, as displacements occur, the unit cell may shift (though the lattice is unchanged) $\mathbb{L} \rightarrow \mathbb{L}\mu_g^{-1}$. Putting the coordinates into crystal normal form dictates these changes. With each shift, the basis coordinates will change: $\mu_g \mathbb{B}_2 = \mu_g \mathbb{B}_1 + \sum \mu_g \vec{\delta_i}$. Because $\{\vec{\delta_i}\}$ is closed with respect to all possible unimodular multiplication, even in the new cell setting a finite path between the starting and ending bases exists, and the next step of the path is always accessible.

Eventually the path terminates and the bases coincide. This process can be repeated for any $\mathbb{B}_2$ of the original unit cell, and, from the perspective of this unit cell, all possible basis relative coordinates are accessible. This process can be repeated also for any different unit cell, and the same results hold, regardless of the unit cell chosen.

\section{Appendix}

Here, we write a complete enumeration of lattice neighbor displacements: $\{\vec{\chi_i}\}$ (42 elements)

$$\pm\left\{\begin{pmatrix}1\\-1\\0\\0\\0\\0\\0\end{pmatrix},\begin{pmatrix}1\\0\\-1\\0\\0\\0\\0\end{pmatrix},\begin{pmatrix}1\\0\\0\\-1\\0\\0\\0\end{pmatrix},\begin{pmatrix}1\\0\\0\\0\\1\\0\\0\end{pmatrix},\begin{pmatrix}1\\0\\0\\0\\0\\1\\0\end{pmatrix},\begin{pmatrix}1\\0\\0\\0\\0\\0\\1\end{pmatrix},\begin{pmatrix}0\\1\\-1\\0\\0\\0\\0\end{pmatrix},\begin{pmatrix}0\\1\\0\\-1\\0\\0\\0\end{pmatrix},\begin{pmatrix}0\\1\\0\\0\\1\\0\\0\end{pmatrix},\begin{pmatrix}0\\1\\0\\0\\0\\1\\0\end{pmatrix},\begin{pmatrix}0\\1\\0\\0\\0\\0\\1\end{pmatrix},\begin{pmatrix}0\\0\\1\\-1\\0\\0\\0\end{pmatrix},\begin{pmatrix}0\\0\\1\\0\\1\\0\\0\end{pmatrix},\begin{pmatrix}0\\0\\1\\0\\0\\1\\0\end{pmatrix},\begin{pmatrix}0\\0\\1\\0\\0\\0\\1\end{pmatrix},\begin{pmatrix}0\\0\\0\\1\\1\\0\\0\end{pmatrix},\begin{pmatrix}0\\0\\0\\1\\0\\1\\0\end{pmatrix},\begin{pmatrix}0\\0\\0\\1\\0\\0\\1\end{pmatrix},\begin{pmatrix}0\\0\\0\\0\\1\\-1\\0\end{pmatrix},\begin{pmatrix}0\\0\\0\\0\\1\\0\\-1\end{pmatrix},\begin{pmatrix}0\\0\\0\\0\\0\\1\\-1\end{pmatrix}\right\}$$

Here, we write a complete enumeration of crystal basis neighbor displacements: $\{\vec{\delta_i}\}$. These displacements can be applied to any basis atom of the crystal.

$$\pm \left\{
\begin{pmatrix}
1 \\ 0 \\ 0
\end{pmatrix}
\begin{pmatrix}
0 \\ 1 \\ 0
\end{pmatrix}
\begin{pmatrix}
0 \\ 0 \\ 1
\end{pmatrix}
\begin{pmatrix}
1 \\ 1 \\ 1
\end{pmatrix}
\right\}$$

Here, we calculate all the neighbors of an example crystal in crystal normal form:

The original crystal in crystal normal form: \\
6	6	15	16	4	19	20	2	10	10

The lattice neighbors of the original crystal: \\
6	6	15	16	4	18	21	2	10	10 \\
6	6	15	16	4	19	20	10	2	10 \\
6	6	15	16	3	19	21	2	10	10 \\
6	6	15	16	5	19	19	2	10	10 \\
6	6	15	16	3	20	20	2	10	10 \\
6	6	15	16	5	18	20	2	10	10 \\
6	6	15	15	4	19	19	0	8	10 \\
6	6	15	17	4	19	21	2	10	10 \\
6	6	15	15	4	18	20	0	8	10 \\
6	6	15	17	4	20	20	2	10	10 \\
6	6	15	15	3	19	20	0	8	10 \\
6	6	15	17	5	19	20	2	10	10 \\
6	6	14	16	4	19	19	2	10	10 \\
6	6	16	16	4	19	21	0	8	10 \\
6	6	14	16	4	18	20	2	10	10 \\
6	6	16	16	4	20	20	0	8	10 \\
6	6	14	16	3	19	20	2	10	10 \\
6	6	16	16	5	19	20	0	8	10 \\
6	6	14	17	4	19	20	2	10	10 \\
6	6	15	16	4	19	20	0	8	10 \\
5	6	15	16	4	19	19	10	2	10 \\
6	7	15	16	4	19	21	2	10	10 \\
5	6	15	16	4	20	18	10	2	10 \\
6	7	15	16	4	20	20	2	10	10 \\
5	6	15	16	3	20	19	10	2	10 \\
6	7	15	16	5	19	20	2	10	10 \\
5	6	15	17	4	20	19	10	2	10 \\
6	7	15	15	4	19	20	2	10	10 \\
5	6	16	16	4	19	20	0	8	10 \\
6	7	14	16	4	19	20	2	10	10 \\
5	6	15	16	4	19	19	2	10	10 \\
6	7	15	16	4	21	19	10	2	10 \\
5	6	15	16	4	18	20	2	10	10 \\
6	7	15	16	4	20	20	10	2	10 \\
5	6	15	16	3	19	20	2	10	10 \\
6	7	15	16	5	20	19	10	2	10 \\
5	6	15	17	4	19	20	2	10	10 \\
6	7	15	15	4	19	20	0	8	10 \\
5	6	16	16	4	19	20	2	10	10 \\
6	7	14	16	4	20	19	10	2	10 \\
5	7	15	16	4	19	20	2	10	10 \\
5	7	15	16	4	20	19	10	2	10 \\

The basis neighbors of the original crystal: \\
6	6	15	16	4	19	20	1	9	9 \\
6	6	15	16	4	19	20	3	11	11 \\
6	6	15	16	4	19	20	2	10	9 \\
6	6	15	16	4	19	20	2	10	11 \\
6	6	15	16	4	19	20	2	9	10 \\
6	6	15	16	4	19	20	2	11	10 \\
6	6	15	16	4	19	20	1	10	10 \\
6	6	15	16	4	19	20	3	10	10

\section{References}
\bibliographystyle{plain}
\bibliography{main}

\begin{thebibliography}{1}

\bibitem{bain1924nature}
Edgar~C Bain and NY~Dunkirk.
\newblock The nature of martensite.
\newblock {\em trans. AIME}, 70(1):25--47, 1924.

\bibitem{barrett1963crystal}
CS~Barrett, P~Cucka, and KJAC Haefner.
\newblock The crystal structure of antimony at 4.2, 78 and 298 k.
\newblock {\em Acta Crystallographica}, 16(6):451--453, 1963.

\bibitem{conway1992low}
John~Horton Conway and Neil~JA Sloane.
\newblock Low-dimensional lattices. vi. voronoi reduction of three-dimensional
  lattices.
\newblock {\em Proceedings of the Royal Society of London. Series A:
  Mathematical and Physical Sciences}, 436(1896):55--68, 1992.

\bibitem{doi:10.1137/S1064827502407627}
James Demmel and Yozo Hida.
\newblock Accurate and efficient floating point summation.
\newblock {\em SIAM Journal on Scientific Computing}, 25(4):1214--1248, 2004.

\bibitem{ekman1998ab}
Mathias Ekman, Babak Sadigh, Kristin Einarsdotter, and Peter Blaha.
\newblock Ab initio study of the martensitic bcc-hcp transformation in iron.
\newblock {\em Physical Review B}, 58(9):5296, 1998.

\bibitem{koumatos2016optimality}
Konstantinos Koumatos and Anton Muehlemann.
\newblock Optimality of general lattice transformations with applications to
  the bain strain in steel.
\newblock {\em Proceedings of the Royal Society A: Mathematical, Physical and
  Engineering Sciences}, 472(2188):20150865, 2016.

\bibitem{sheppard2012generalized}
Daniel Sheppard, Penghao Xiao, William Chemelewski, Duane~D Johnson, and Graeme
  Henkelman.
\newblock A generalized solid-state nudged elastic band method.
\newblock {\em The Journal of chemical physics}, 136(7):074103, 2012.

\bibitem{stokes2002procedure}
Harold~T Stokes and Dorian~M Hatch.
\newblock Procedure for obtaining microscopic mechanisms of reconstructive
  phase transitions in crystalline solids.
\newblock {\em Physical Review B}, 65(14):144114, 2002.

\bibitem{yamanaka2000x}
T~Yamanaka, R~Kurashima, and J~Mimaki.
\newblock X-ray diffraction study of bond character of rutile-type sio2, geo2
  and sno2.
\newblock {\em Zeitschrift f{\"u}r Kristallographie-Crystalline Materials},
  215(7):424--428, 2000.

\end{thebibliography}

\end{document}